\documentclass[%
 reprint, 
 amsmath,amssymb,
 aps,
 nofootinbib
 prb,
 superscriptaddress
 ]{revtex4-1}
\usepackage{scrextend}
\usepackage{graphicx}
\usepackage{dcolumn}
\usepackage{bm}
\usepackage{hhline}
\usepackage{tabularx}
\usepackage[utf8]{inputenc}
\usepackage[english]{babel}
\usepackage[bottom]{footmisc}
\usepackage{color}

\begin{document}

\preprint{APS/123-QED}

\title{First-principles investigations of orthorhombic-cubic phase transition and its effect on thermoelectric properties in
cobalt-based ternary alloys.}
\author{Sapna~Singh}\thanks{S.S. and M.Z. contributed equally to this work.} 
\affiliation{Indian Institute of Technology Roorkee, Department of Chemistry, Roorkee 247667, Uttarakhand, India}
\author{Mohd~Zeeshan}\thanks{S.S. and M.Z. contributed equally to this work.}
\affiliation{Indian Institute of Technology Roorkee, Department of Chemistry, Roorkee 247667, Uttarakhand, India}
\author{Jeroen~van~den~Brink}
\affiliation{Institute for Theoretical Solid State Physics, IFW Dresden, Helmholtzstrasse 20, 01069 Dresden, Germany}
\author{Hem~C.~Kandpal}\email{Corresponding author: hem12fcy[at]iitr.ac.in}
\affiliation{Indian Institute of Technology Roorkee, Department of Chemistry, Roorkee 247667, Uttarakhand, India}

\date{\today}
             
\begin{abstract} 
We screened six cobalt-based 18-VEC systems CoVSi, CoNbSi, CoTaSi (Si-group) and CoVGe, CoNbGe, CoTaGe (Ge-group) by the first-principles approach, with the motivation of stabilizing these orthorhombic phases into the cubic symmetry -- favorable for thermoelectrics. Remarkably, it was found that the Ge-group is energetically more favorable in the cubic symmetry than the
hitherto orthorhombic phase. We account the cubic ground state of the Si-group to the interplay of internal pressure and covalent interactions. The principle of covalent interactions will provide an insight and could be vital in speeding the search of missing cubic half-Heusler alloys. Meanwhile, the calculated transport properties of all the systems on \textit{p}-type doping, except CoVSi, are more promising than the well-known CoTiSb. We also provide conservative estimates of the figure of merit, exceeding the CoTiSb. Based on our findings, we suggest possible new phases of ternary compounds for thermoelectric applications. 

\begin{description}
\item[Usage]
Secondary publications and information retrieval purposes.
\item[PACS numbers]
May be entered using the \verb+\pacs{#1}+ command.
\item[Structure]
You may use the \texttt{description} environment to structure your abstract;
use the optional argument of the \verb+\item+ command to give the category of each item. 
\end{description}
\end{abstract}

\pacs{Valid PACS appear here}
\maketitle

\section{Introduction} 
First-principles investigations have been helpful in designing new materials with the benefits of screening a large number of known and unknown materials in different symmetries \cite{Bennett13, Roy12}. This way one can tune the materials for desirable properties as materials behave differently in different symmetries. Certainly, the stability of the desired symmetry is of key importance. Nonetheless, by targeting a particular class of materials, the properties of interest can be obtained by optimizing the materials in a suitable geometry. Our target class of materials is novel half-Heusler (hH) alloys which crystallize in cubic \textit{F$\bar{4}$3m} MgAgAs structure type \cite{Offernes07}. The other competing symmetries are hexagonal \textit{P6$_3$/mmc} ZrBeSi structure type \cite{Bennett12} and orthorhombic \textit{Pnma} MgSrSi structure type \cite{Conrad05}. The cubic symmetry has been exploited most as they offer numerous tunable properties based on their valence electron count (VEC) \cite{Pierre97}. Likewise, they have diverse potential applications such as in spintronics \cite{Casper12}, superconductors \cite{Xiao18}, and thermoelectrics \cite{Zhu15}, to name a few. 

As potential thermoelectric (TE) materials, 18-VEC cubic hH alloys have been extensively studied over the past two decades in search of better prospects of converting the waste heat into electricity, governed by a dimensionless quantity called the figure of merit (\textit{ZT}) \cite{Xie12, Chen13}. Despite the significant advances, the highest \textit{ZT} achieved in hH alloys is 1.6 \cite{Yu17}. However, to compete with the conventional methods of power generation, a \textit{ZT} $\sim$ 3-4 is much needed \cite{He17}. Therefore, along with improving the existing materials, the key research would be developing new hH alloys. The current investigations in hH alloys are centered on 18-VEC cubic phases. Can competing orthorhombic or hexagonal analogue of hH alloys be exploited for thermoelectrics? The most probable answer -- No. This is because of the metallic behavior of orthorhombic and hexagonal hH phases whereas semiconductors are the best starting materials for the theremoelectrics, as established long back by Ioffe \cite{Vedernikov98}. Mostly, cobalt-based 18-VEC cubic hH alloys are semiconducting in accordance with the Slater-Pauling rule \cite{Fecher06}. Some of the well-explored cobalt-based hH alloys as
potential TE materials are CoTiSb, CoZrSb, CoHfSb \cite{Sekimoto05}, and CoNbSn \cite{He16} whereas the TE properties of CoVSn, CoTaSn, CoMoIn, and CoWIn were recently proposed \cite{Zeeshan17}. However, the cobalt-based 18-VEC systems having main group element Si and Ge have not been explored for thermoelectricity as they crystallize in orthorhombic symmetry.

Can these 18-VEC orthorhombic cobalt-based systems be stabilized in cubic symmetry? Historically, the most amazing phase transitions known to mankind are controlled by temperature and pressure. Nature offers the fascinating example of the formation of diamond in the earth crust by intense temperature and pressure. In laboratories, a number of phase transitions are reported which are governed by temperature, pressure, and in some cases by varying synthetic techniques; a detailed review can be found in the literature \cite{Trigunayat91}. Particularly, in hH alloys, the hydrostatic pressure induced cubic-hexagonal phase transition was reported in FeVSb and CoVSb \cite{Noda79}. In a special derivative of hH alloys, Nowotny-Juza phases, the role of hydrostatic and internal pressure was emphasized in cubic-hexagonal phase transition by first-principles calculations \cite{Chopra18}. 

In a recent first-principles based search of new TE prospects, 18-VEC orthorhombic systems CoVSi, CoNbSi, and CoTaSi were doped with Sn \cite{Samsonidze17}. Beyond a certain doping level, the cubic phase was found to be the ground state, i.e., favorable for thermoelectrics. Is it merely doping effect or pressure has some role to play? As one starts replacing Si by Sn in the unit cell, a relative expansion of the crystal lattice occurs which corresponds to the building of negative internal pressure within the system. This suggests that the internal pressure might be the governing force behind the phase transition. A number of studies have been reported where internal pressure has a profound impact in controlling the phase transitions \cite{Huon17, Horiuchi03, Fratini08, Dhital17}. All in all, both hydrostatic and internal pressure seems to play an important role in phase transitions in hH alloys. Such pressure based phase transitions motivated us to investigate the orthorhombic-cubic phase transition and its impact on transport properties in 18-VEC cobalt-based ternary systems. 

In this paper, utilizing \textit{ab initio} approach, we investigate the role of pressure (hydrostatic/internal) in governing the orthorhombic-cubic phase transition in six 18-VEC cobalt-based systems; CoVSi, CoNbSi, CoTaSi, CoVGe, CoNbGe, and CoTaGe. Also, we consider the hexagonal symmetry as the intermediate symmetry between orthorhombic-cubic phase transition. The reason is the existence of some high pressure hexagonal phases of hH alloys. Further, the impact of phase transition on transport properties is studied -- the prime motive of the work. In order to completely understand the phase transition and its impact on transport properties, we also include systems based on Sn, i.e., CoVSn, CoNbSn, and CoTaSn. The inclusion of these systems is quite obvious as Si and Ge are followed by Sn in carbon-family. Furthermore, these systems are reported in the cubic symmetry and will  help in understanding the orthorhombic-cubic phase transition. 

The paper is organized into four sections. Section II briefly describes the computational tools utilized
in this paper. Section III elucidates the results which comprise the crystal structure optimization, dynamic
stability of the systems in considered symmetries, and electrical and thermal transport properties in cubic
symmetry. Finally, the important findings and remarks are mentioned in Sec. IV. 


\section{Computational Details}
We used a combination of two different first-principles density functional theory (DFT) codes: the full-potential linear augmented plane wave method (FLAPW) \cite{Singh06} implemented in Wien2k \cite{Blaha01} and the plane-wave pseudopotential approach implemented in Quantum Espresso package \cite{Giannozzi09}. The former has been used to obtain equilibrium lattice constants, electronic structure, and transport properties, and the latter to confirm the structure stability by determining the phonon spectrum. 

The FLAPW calculations were performed using a modified Perdew-Burke-Ernzerhof (PBEsol correlation) \cite{Perdew08} implementation of the generalized gradient approximation (GGA). For all the calculations, the scalar relativistic approximation was used. The muffin-tin radii (RMTs) were taken in the range 1.86-2.4 Bohr radii for all the atoms. RMT {$\times$} kmax = 7 was set as the plane wave cutoff. The self-consistent calculations were employed using 125000 \textit{k}-points in the full Brillouin zone. The energy and charge convergence criterion was set to 10$^{-6}$ Ry and 10$^{-5}$ e, respectively. 

The electrical transport properties have been calculated using the Boltzmann theory \cite{Allen} and relaxation time approximation as implemented in the Boltztrap code \cite{Madsen06}. The Boltztrap code utilizes input from Wien2k code. The electrical conductivity and power factor are calculated with respect to time relaxation, \textit{$\tau$}; the Seebeck coefficient is independent of \textit{$\tau$}. The relaxation time was calculated by fitting the available experimental data with theoretical data. We have used this approach in evaluating the electronic transport properties of our systems.  

In the plane-wave pseudopotential approach, we used scalar-relativistic, norm-conserving pseudopotentials for a plane-wave cutoff energy of 100~Ry. The exchange-correlation energy functional was evaluated within the GGA, using the Perdew-Burke-Ernzerhof parametrization \cite{Perdew96}, and the Brillouin zone was sampled with a 20$\times$20$\times$20 mesh of Monkhorst-Pack \textit{k}-points. The calculations were performed on a 2$\times$2$\times$2 \textit{q}-mesh in the phonon Brillouin zone.  

The lattice thermal conductivity was obtained by solving linearized Boltzmann transport equation (BTE) within the single-mode relaxation time approximation (SMA) \cite{Ziman1960} using thermal2 \cite{thermal2} code as implemented in the Quantum Espresso package. We employed generalized gradient approximation (GGA) given by Perdew-Burke-Ernzerhof (PBE) \cite{Perdew96} for exchange-correlation functional. We have chosen Troullier-Martins norm-conserving pseudopotentials from the Quantum Espresso webpage \cite{QEweb}. An energy cutoff of 100~eV was used for the plane-wave basis set and Brillouin zone integration was performed on a Monkhorst-Pack 20$\times$20$\times$20.

\section{Results}
This section briefly discusses the plausible crystal structures of hH alloys and their structural optimization. Followed by their dynamical stability and finally the analysis of electrical and thermal transport properties of stable cubic semiconducting systems.

\subsection{Crystal Structure}
\begin{figure}
\centering\includegraphics[scale=0.35]{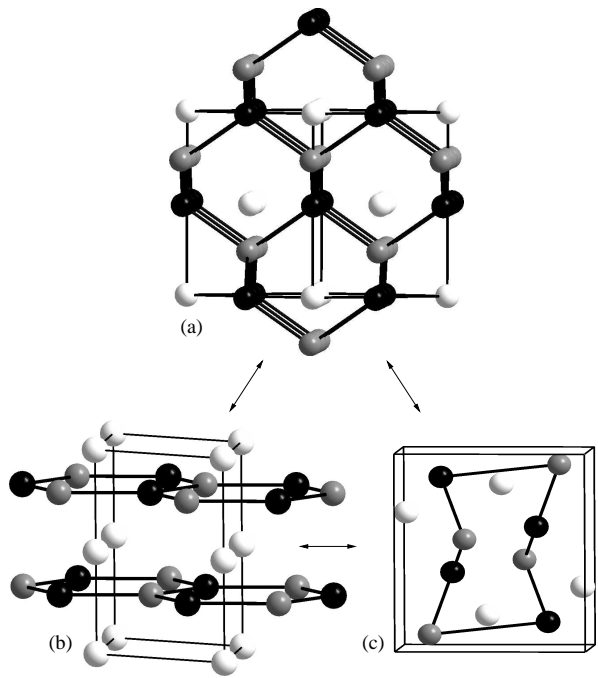}
\caption{The possible crystal structures of Co\textit{YZ} systems (\textit{Y} = V, Nb, Ta, and \textit{Z} = Si, Ge, Sn)
in (a) cubic \textit{$F\bar{4}3m$}, (b) hexagonal \textit{P6$_3$/mmc}, and (c) orthorhombic \textit{Pnma} symmetry.}
\label{crystal_structure}
\end{figure}

The crystal structure of hH alloys, XYZ, in three plausible geometries viz. cubic \textit{$F\bar{4}3m$}, hexagonal \textit{P6$_3$/mmc}, and orthorhombic $Pnma$ is shown in Fig.~\ref{crystal_structure}. The XZ forms a puckered octagonal framework in orthorhombic symmetry whereas a planar hexagonal framework in the \textit{P6$_3$/mmc} symmetry. A somewhat similar hexagonal framework can be seen in the cubic symmetry. This suggests that the desired orthorhombic-cubic phase transition might be intermediated through hexagonal symmetry. Very plausible, the stabilization of the orthorhombic phase into the hexagonal one. However, it may not be relatively easier to interconvert these structures as it may require a great deal of pressure. These crystal structures are well explored and can be seen in detail in Ref. \cite{Chopra18, Conrad05}. 

\subsection{Structural Optimization}
Throughout this paper, we use the term Si-group to refer to CoVSi, CoNbSi, CoTaSi; Ge-group to refer to CoVGe, CoNbGe, CoTaGe, and Sn-group to refer to CoVSn, CoNbSn, CoTaSn. The ground state properties of all the symmetries were obtained using GGA-PBEsol implemented in Wien2k. The PBEsol approximation was chosen as it is one of most accurate approximation for calculating the ground state properties \cite{He14}. Fitted with Birch-Murnaghan equation \cite{Birch47}, the optimized lattice parameters were obtained by minimizing the total energy as a function of variable structural parameters in the respective symmetry. The variable parameter in case of cubic symmetry was volume, in case of hexagonal \textit{P6$_3$/mmc} were volume and c/a; additionally variable parameters b/a and atomic positions were also optimized in case of orthorhombic symmetry. Thuswise optimized lattice parameters in different symmetries are listed in Table~\ref{optimization}. The calculated lattice parameters are underestimated as compared to the experimental values. The underestimate of lattice parameters is up to 1\% for $a$ and 1.5\% for $b$ and $c$ in orthorhombic symmetry, along with some higher discrepancies. The discrepancy in the case of CoVSi and CoVGe is around 3\% in lattice parameter $b$ and 2\% in $c$-parameter. Except for CoVSn, the discrepancy in the case of cubic Sn-group was less than 1\%, which is quite acceptable. The higher discrepancy (4\%) in case of CoVSn is acceptable as the reported sample was partially disordered. Moreover, the other theoretical works on CoVSn have also reported the similar discrepancy \cite{Hichour12, Ameri13}.

The trend of lattice parameter in all the symmetries is uniform. The lattice parameter $a$, first, increases from the first member to second and then decreases marginally in the third member for Si-, Ge- and Sn-group in either symmetry. The slight decrease along the third member of either group can be attributed to the lanthanide contraction on post lanthanide elements \cite{Housecroft04, Cotton88}. The b/a ratio is almost constant at 0.59 whereas the c/a ratio ranges 1.13--1.15 in orthorhombic symmetry. The c/a ratio in case of hexagonal symmetry ranges from 1.20 to 1.40. 

\begin{table*}
\caption{The optimized cell parameters (in \AA) of Co\textit{YZ} systems (\textit{Y} = V, Nb, Ta, and \textit {Z} = Si, Ge, Sn ) in cubic
\textit{$F\bar{4}3m$}, hexagonal \textit{P6$_3$/mmc}, and orthorhombic \textit{Pnma} symmetry. The band gap values (in eV) in cubic
\textit{$F\bar{4}3m$} symmetry are also listed whereas the systems in hexagonal and orthorhombic symmetry are metallic. The last column
lists the reference of the reported symmetry.}
\setlength{\arrayrulewidth}{0.5pt}
\begin{tabularx}{\textwidth}{c @{\extracolsep{\fill}} cccccccccc}
\hline
\hline
		&\multicolumn{2}{c}{\textit{$F\bar{4}3m$}}	&\multicolumn{2}{c}{\textit{P6$_3$/mmc}} &\multicolumn{3}{c}{\textit{Pnma}} &\multicolumn{1}{c}{Ref.}	\\ \hline		
System		&a		&E$_g$				&a		    &c/a			 &a	      	&b/a	&c/a  	     		\\ \hline
CoVSi		&5.3679		&0.544				&4.0141		&1.23			 &5.8888	&0.59	&1.14 	&\textit{Pnma} \cite{Jeitschko69}		 		\\
CoNbSi		&5.5811 	&0.845				&4.0437		&1.38			 &6.1873	&0.59	&1.13 	&\textit{Pnma} \cite{Jeitschko69}		 		 \\
CoTaSi		&5.5807		&1.226				&4.0214		&1.40			 &6.1575	&0.59	&1.13 	&\textit{Pnma} \cite{Jeitschko69}		   		 \\
CoVGe		&5.4521		&0.536				&4.0762		&1.23			 &5.9768	&0.60	&1.15 	&\textit{Pnma} \cite{Jeitschko69}		   		 \\
CoNbGe		&5.6476		&1.105				&4.1471		&1.38			 &6.2541	&0.59	&1.13 	&\textit{Pnma} \cite{Jeitschko69}				 \\
CoTaGe		&5.6441		&1.187				&4.1293		&1.34			 &6.2336	&0.59	&1.13	&\textit{Pnma} \cite{Jeitschko69}			 	 \\
CoVSn		&5.7338		&0.759				&4.3305		&1.20			 &6.3675	&0.60	&1.15	&\textit{$F\bar{4}3m$} \cite{Lue01}			 	 \\
CoNbSn		&5.8983		&1.103				&4.4268		&1.24			 &6.6309	&0.59	&1.13	&\textit{$F\bar{4}3m$} \cite{Buffon16}			  	 \\
CoTaSn		&5.8903		&1.107				&4.4136		&1.25			 &6.6237	&0.59	&1.13 	&\textit{$F\bar{4}3m$} \cite{Zakutayev13}				 \\ 
\hline
\hline
\end{tabularx}
\label{optimization}
\end{table*}

\begin{figure*}
\centering\includegraphics[scale=0.6]{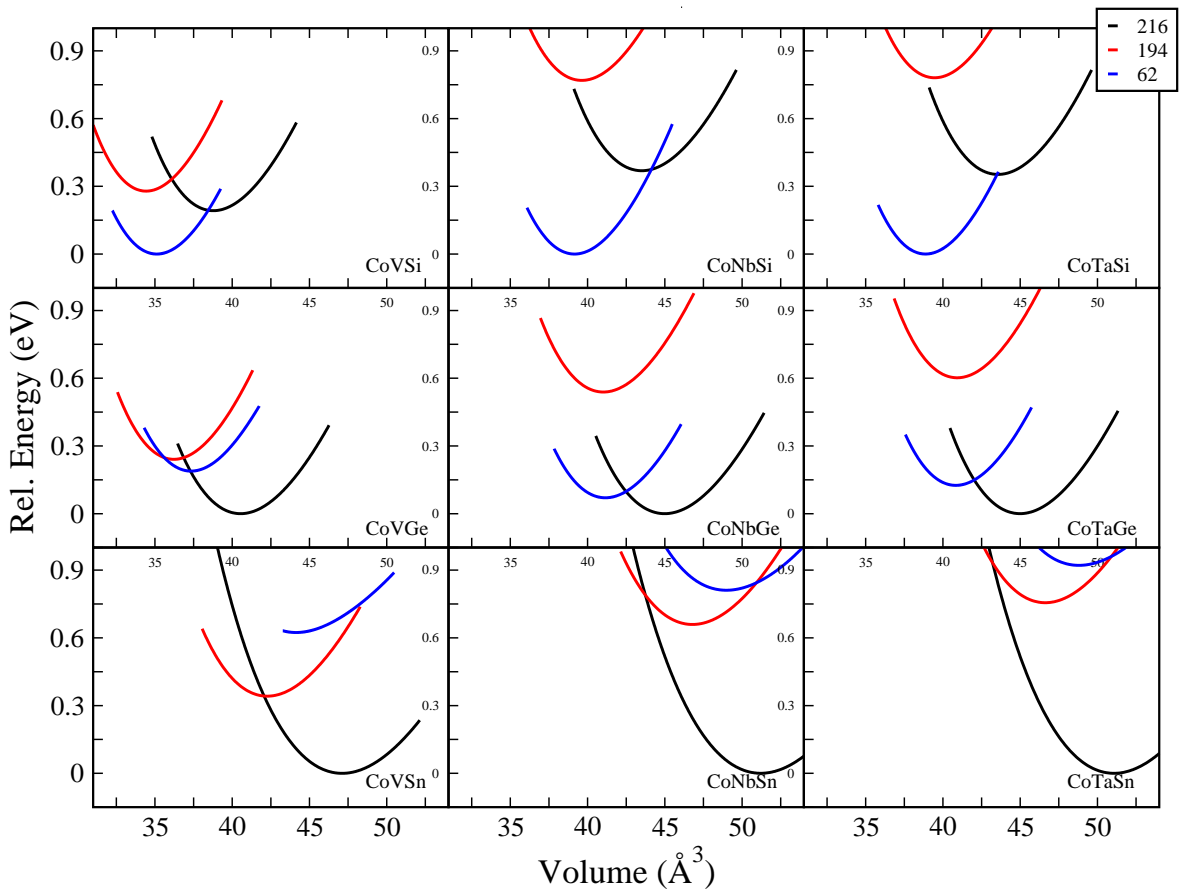}
\caption{The relative energy as a function of the volume of Co\textit{YZ} systems (\textit{Y} = V, Nb, Ta, and \textit{Z} = Si, Ge, Sn)
in cubic \textit{$F\bar{4}3m$}, hexagonal \textit{P6$_3$/mmc}, and orthorhombic \textit{Pnma} symmetry. The ground state symmetry
in each case is set as zero on the vertical axes.}
\label{plot}
\end{figure*}

Focussing on another aspect of Table~\ref{optimization}, i.e., band gap, all the systems were found to be metallic in both orthorhombic and hexagonal symmetry. Nevertheless, as expected, all the systems turn out to be semiconducting in the cubic symmetry, i.e., favorable for thermoelectrics. Now, it becomes important to see whether the change in symmetry from orthorhombic to cubic is possible or not. For this, the energy as a function of the volume of all nine systems in three plausible geometries is shown in Fig.~\ref{plot}. None of the systems could be stabilized in the anticipated hexagonal \textit{P6$_3$/mmc} symmetry. On the other hand, remarkably, we found that the previously known orthorhombic Ge-group is energetically more favorable in the cubic symmetry. This could be interesting from the thermoelectrics perspective as none of the Ge-based 18-VEC hH alloys, to the best of our knowledge, are reported in cubic symmetry. If stabilized in the cubic symmetry, there might be chances that these Ge-based hH alloys exhibit competitive figure of merit as those of existing ones. 

Before discussing ahead, we would like to apprehend why the Ge-group was previously reported in orthorhombic symmetry. Experiments in Ge-group were performed long back in 1969 by Jeitschko et al \cite{Jeitschko69}. Since then, there has been no report in these systems either in orthorhombic or cubic symmetry. In our opinion, the energy difference between the cubic and orthorhombic symmetry of the Ge-group might have been the reason behind the existence of these alloys in orthorhombic symmetry. The calculated energy difference between the cubic and orthorhombic symmetry is 0.18, 0.07, and 0.12 eV for CoVGe, CoNbGe, and CoTaGe, respectively, Fig.~\ref{plot}. Considering the room temperature energy of 0.025 eV, it seems that the slight activation energy in the form of temperature or pressure by the synthetic method utilized might have stabilized these structures in orthorhombic symmetry. Recently, in a special derivative of hH alloys, i.e., Nowotny-Juza phases, the cubic phase of LiZnSb was synthesized under the ambient conditions by low-temperature solution phase method despite the long existing hexagonal phase reported by solid-state techniques \cite{White16}. Thereupon, in a theoretical work, it was attributed to the small energy difference between the cubic and hexagonal phase which might have led to the high-temperature hexagonal and low-temperature cubic phase of LiZnSb \cite{Chopra18}. 

Based on such investigations, the proposed cubic symmetry of the Ge-group is not completely surprising and experimentally unrealizable. Keeping in mind the recent success of solution phase method in case of LiZnSb, we believe that the considered systems could be synthesized in cubic symmetry by low-temperature solution phase method; if not by solid-state techniques. Coming back to the optimization plot of Co\textit{YZ} systems, we analyze different aspects in order to understand the varying trends in these systems. The ground state of the Si-group was found to be orthorhombic in accordance with the reported structures. On going down the first column from CoVSi to CoVGe, the cubic symmetry becomes the ground state whereas the orthorhombic goes to the higher energy state, Fig.~\ref{plot}. The energy gap between the cubic ground state and the orthorhombic state becomes even higher in case of CoVSn. A similar trend is observed in the case of second and third column, i.e., going from CoNbSi to CoNbSn and CoTaSi to CoTaSn. This can be attributed to the increasing size of the atom as one moves from Si to Sn. 

The increasing atomic size leads to the relative expansion of the unit cell volume. Such expansion corresponds to the building of negative internal pressure within the crystal lattice which might be the driving force behind these phase transitions. Now, this pressure can also be achieved by isoelectronic doping of Si by Ge or Sn. Beyond a certain doping level, the cubic phase is likely to dominate. This is what Samsonidze et al. \cite{Samsonidze17} observed in the case of Si-group. They doped CoVSi, CoNbSi, and CoTaSi by Sn until they obtained the stable cubic symmetry. Here, we stress that it is more important to understand the underlying phenomenon behind such phase transitions and would like to propose that not only Sn but Ge doping can also be helpful in obtaining the cubic symmetry. Moreover, it would be relatively easier to replace Si by Ge as compared to Sn.  

Meanwhile, another way of introducing the internal pressure within the system is the introduction of the nonreactive entities such as hydrogen or helium in the vacant sites. It might be challenging to introduce the nonreactive species into the crystal lattice, however, the solution phase chemical route could be a favorable approach to achieve the hydrogen insertion. Besides, the vacant sites in the hH structures may facilitate the hydrogen insertion. Moreover, optimistic of the experimental evidence of hydrogen absorption in the crystal lattice of intermetallics, we do not see this as completely unachievable. Soubeyroux et al. studied the hydrides of ternary ZrCo(V$_{1-x}$Cr$_x$) \cite{Soubeyroux99}. At 100 $^0$C, it was found that the reversible hydrogen capacity was 2.6 H/f.u. In another work, Zr(Cr$_{1-x}$Co$_x$)$_2$ and its hydrides were studied \cite{Hirosawa83}. It was found that the hydrogen capacity of the alloy, at room temperature, was 3.6 H atoms/f.u. In a more unique approach, such pressure can be achieved by introducing the epitaxial strain in the grown thin films of these alloys. Some of the epitaxial thin films of half-Heusler \cite{Wen15}, full Heusler \cite{Takamura14}, and quaternary Heusler alloy \cite{Bainsla17} were grown on MgO substrate. To conclude thus far, the cubic symmetry is the optimized ground state of Ge-group whereas orthorhombic Si-group can be stabilized in the cubic symmetry by isoelectronic doping, the introduction of nonreactive entities in the crystal lattice, or in epitaxial strained thin films.

Having discussed the trend along the column, next, we analyze the optimization plot along the row, i.e., from CoVSi to CoTaSi and so on. As seen along the column, the increase in the unit cell volume led to the building of negative internal pressure and thereby change in orthorhombic-cubic symmetry from CoVSi to CoVGe. We expect a similar orthorhombic-cubic symmetry change along the row. Contrary to our expectations, no such trend was observed along the first row on going from CoVSi to CoNbSi to CoTaSi. What is surprising is the fact that the cubic symmetry does not become energetically favorable even at CoTaSi, despite the significant change in volume as compared along the first column. This as well highlights an important point that the doping of CoVSi/CoNbSi/CoTaSi at V-position by Nb or Ta will not result into the cubic symmetry, as occurred by the Ge/Sn-doping at the Si-position. What is so special along the column that misses along the row? Does it have something to do with the bonding picture in the hH alloys? Thus, to gain an insight into this, we take CoTiSb as a case to understand the bonding picture in cubic hH alloys. We choose CoTiSb as it is the most explored cobalt-based hH alloy and in the later section, we also corroborate our calculated transport properties with respect to CoTiSb.

Nonetheless, the chemical bonding picture treats CoTiSb as Ti$^{4+}$ [CoSb]$^{4-}$, or in general Y$^{n+}$ [XZ]$^{n-}$ \cite{Graf11}. Where CoSb forms the covalent framework, the most electropositive Ti donates its electrons to the [CoSb] bond.
In the considered systems, going along the row from CoVSi to CoTaSi, the covalent framework remains intact, i.e., [CoSi], whereas the change in the electropositive atom from V to Ta does not affect the covalent framework much. However, if we move down the column from CoVSi to CoVSn, the covalent framework changes at every step, i.e., [CoSi], [CoGe], and [CoSn]. It is easy to notice that the [CoSn] has the highest degree of covalent interactions whereas [CoSi] has the least. This change in the covalent character was missing along the row. This suggests that the more covalent interactions favor the cubic symmetry in these alloys. As a result, the Si-group stabilizes in orthorhombic symmetry whereas the Ge-group and Sn-group stabilize in cubic symmetry; on account of increased covalent interactions. Based on these observations, we believe that the introduction of covalent interactions within the crystal lattice could be vital in speeding the search of missing cubic hH alloys. 

As stated above, the doping in CoVSi/CoNbSi/CoTaSi at V-position by Nb or Ta will not result into the cubic symmetry, as happened by the Ge/Sn-doping at the Si-position. Here, the covalent interactions driven orthorhombic-cubic phase transition opens up more possibilities of doping. In order to synthesize the cubic structures of Si-group, in addition to the Ge/Sn-doping
at Si-position, one can tune the composition in such a way that maximizes the covalent interactions. As a guide to experimentalists, we suggest more complicated compositions such as Co$_{1-x}$FeVSi$_{1-y}$Sb or Co$_{1-x}$NiVSi$_{1-y}$In in place of parent CoVSi; maintaining the 18-VEC. The dopants suggested are non-toxic, easy to handle, and not very expensive. Most importantly, it will bring the desired covalent interactions required for the stabilization in the cubic symmetry. Also, the complicated compositions facilitate the scattering of phonons thereby minimizing the thermal conductivity. Similarly, the CoNbSi and CoTaSi can also be doped with suitable dopants in order to crystallize in the cubic symmetry.  

We have considered different aspects of the optimization plots, i.e., along the row and column. Needless to say, along the diagonal, a similar kind of behavior can be observed. Going diagonally from CoVSi to CoVGe, the increased covalent interactions
stabilize the cubic symmetry, also preserved in the succeeding CoTaSn. This opens up another possibility of doping, i.e., CoVSi can be doped by Nb at V and Ge at Si-position or by Ta and Sn to stabilize in the cubic symmetry. Likewise, following the another diagonal, the CoTaSi can be stabilized in the cubic symmetry by doping Nb and Ge or V and Sn. The important aspect is to introduce the covalent interactions within the system. The importance of covalent interactions in orthorhombic-cubic symmetry change is appreciable, however, the role of pressure and thus volume cannot be completely overlooked. Not only the covalent character will always stabilize the cubic symmetry in ternary alloys. Had it been the case, almost all the alkali metal-based ternary alloys would have crystallized in cubic symmetry. Therefore, a proper balance between the covalent interactions and volume of the system is needed. An optimum volume is necessarily required for the breathing space of cubic symmetry, in addition to the covalent interactions.  

\begin{figure}
\centering\includegraphics[scale=0.45]{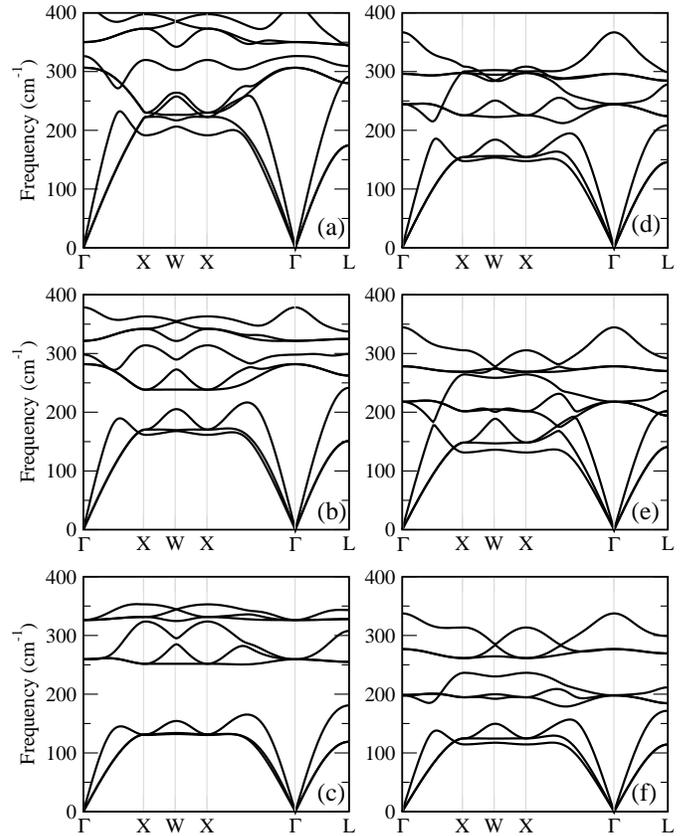}
 \caption{(a-f) The phonon dispersion curves of CoVSi, CoNbSi, CoTaSi, CoVGe, CoNbGe, and CoTaGe, respectively, in the proposed cubic \textit{F$\bar4$3m} symmetry.}
\label{phonon}
\end{figure}

To summarize, we propose that the Ge-group is energetically more favorable in the cubic symmetry whereas the Si-group can be stabilized in the cubic symmetry by different approaches, as discussed. Further, we propose that the covalent interactions and internal pressure can be the governing force in stabilizing the cubic ground state of ternary alloys. Concurrently, we arrived at our aim of stabilizing the 18-VEC orthorhombic hH alloys in the cubic symmetry -- favorable for thermoelectrics. The formation energy of the considered systems in Open Quantum Materials Database (OQMD) further substantiates the stability in the cubic symmetry \cite{Saal13, Kirklin15}. Most importantly, the band gap survived in all the cases. As a next step, we check the dynamical stability of all the six systems in the proposed cubic symmetry -- whether the systems in the proposed symmetry could survive or not. 

\subsection{Structural stability}
To check the dynamical stability of the considered systems in the proposed cubic symmetry, we performed a two-step phonon calculation. This was done by means of Quantum Espresso (QE). First, we tried to match the optimized lattice parameters of Wien2k by QE. The Wien2k optimized lattice parameters were underestimated by QE with a maximum of 2.6\%. Secondly, we calculated the phonon dispersion curves by density functional perturbation theory (DFPT) implemented within the QE. The calculations were performed on a 2$\times$2$\times$2 mesh in the phonon Brillouin zone, and force constants in the real space derived from this input were used to interpolate between q points and to obtain the continuous branches of the phonon band structure. 

Phonons are basically the normal modes or quantum of vibrations in a crystal which helps in governing the stability of a system. For a system to be dynamically stable, it should have only real frequencies and not imaginary \cite{Elliott06, Togo15}. As we are interested in the thermoelectricity, Fig.~\ref{phonon} shows the phonon dispersion frequencies of the six cobalt-based hH alloys in the proposed cubic symmetry. As can be seen from Fig.~\ref{phonon}, all the systems have positive phonon frequencies and can be considered as stable in the cubic symmetry, if synthesized. Now, we are assured that the six cobalt-based systems in the proposed cubic symmetry are semiconducting and dynamically stable, we check their transport properties in the next section. 

\subsection{Transport Properties}
In this section, we calculate the transport properties of six cobalt-based systems, i.e., Si- and Ge-group, in the proposed cubic symmetry. In previous works \cite{Zeeshan17, Zeeshan17_II}, we have already reported the transport properties of Sn-group, therefore, we exclude them for further investigation. The efficiency of a TE material is given by a dimensionless quantity called figure of merit, given by $ZT=S^2\sigma T/\kappa (\kappa = \kappa_e + \kappa_l)$, where S is the Seebeck coefficient, $\sigma$ is the electrical conductivity, and $\kappa$ is the thermal conductivity which comprises electronic and lattice components \cite{Sevincli13, Boona17, Tan17}. Seebeck coefficient, electrical conductivity, and electronic thermal conductivity are calculated by Boltztrap code which relies on the input from the Wien2k code. As $\sigma$ and $\kappa_e$ are relaxation time dependent, which presently cannot be computed by Boltztrap code, we present these quantities under the constant relaxation time approximation (CRTA). The doping effect on the transport parameters is studied under the rigid band approximation (RBA). The CRTA and RBA have been the useful paradigm and widely used in theoretical prediction of the transport properties in a number of systems \cite{Madsen06Boltztrap, Lee11, Chaput05, Jodin04}.

\begin{figure}
\centering\includegraphics[scale=0.55]{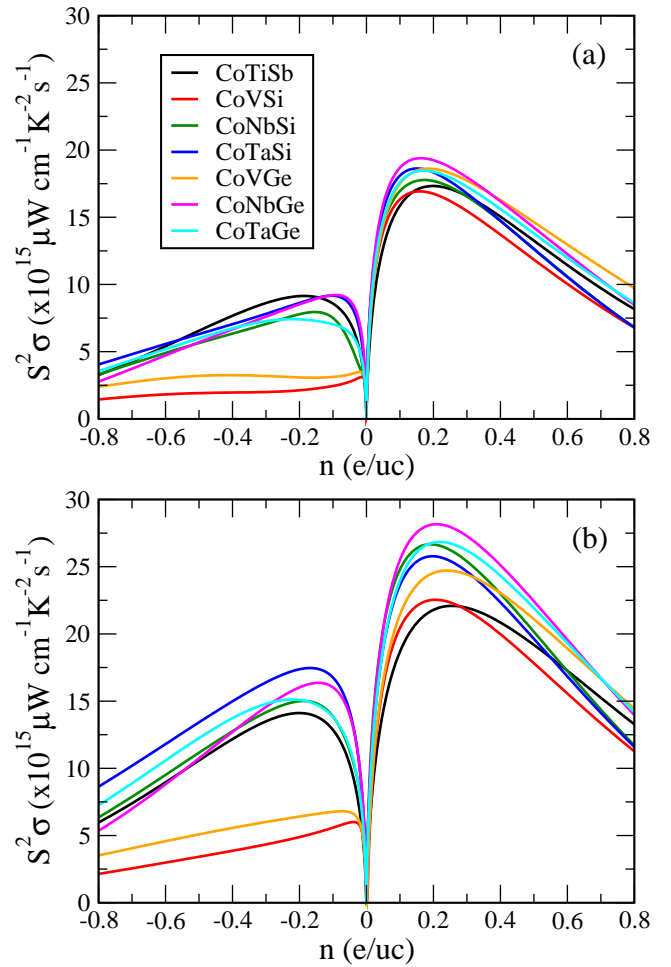}
 \caption{Power factor as a function of doping of Co\textit{YZ} systems (\textit{Y} = V, Nb, Ta, and \textit{Z} = Si, Ge) at (a) 700 K and (b) 1100 K in cubic \textit{F$\bar4$3m} symmetry. The \lq{+\rq} and \lq{--\rq} signs on the horizontal axes represent the hole and electron
 doping, respectively.}
\label{PF}
\end{figure}

The power factor (PF), $S^2 \sigma$, as a function of doping (per unit cell) of the Si- and Ge-group in the cubic symmetry at 700 and 1100 K is shown in Fig.~\ref{PF}. Optimistic of the previous experimental studies \cite{Sekimoto05, Sekimoto06, Sekimoto05_II} and our interest in high-temperature TE applications, we chose to present the PF values at such high-temperatures. For the PF values are more impressive at 1100 K, here onwards, our focus lies on the values at 1100 K. We have included CoTiSb as a reference to corroborate our calculated transport properties. This is because the CoTiSb is the most studied cobalt-based hH alloy in both theory and experiment \cite{Galanakis02, Zhou07, Birkel12, Ouardi12, Wu07, Zhou05}. In a previous work, we found a nice agreement between the calculated and reported PF as a function of temperature for parent CoTiSb. Further, our calculated value of \textit{p}-doped CoTiSb was 20.2 $\mu$W cm$^{-1}$ K$^{-2}$ at 900 K whereas the reported value was 23 $\mu$W cm$^{-1}$ K$^{-2}$ at 850 K \cite{Wu07}. This assures that our calculated transport properties for the proposed systems are quite reliable. In addition, we established therein that a relaxation time of $\tau$ = 10$^{-15}$ s, arrived at by exploiting the experimental electrical conductivity values, is quite an apt number for the cobalt-based systems. In the present work, we have utilized this constant number for evaluating the PF and thus the \textit{ZT} values.

As Seebeck coefficient decreases and electrical conductivity increases with doping, the trend of PF is first increasing and then decreasing gradually at higher doping levels for either doping type. It can be seen from Fig.~\ref{PF} that all the systems have better PF on \textit{p}-type doping as compared to the \textit{n}-type. Since the Seebeck and electrical conductivity follows the opposite trend on doping, one should strike a right balance between the two and choose such an optimal doping level which yields the maximum PF. Generally, such a doping level is found near the band edge which is the case with all the proposed systems under study. The prediction of optimal doping levels enables the experimentalists to narrow down the window for targeting new compositions. 

\begin{table*}
\caption{The calculated optimal doping levels and the corresponding Seebeck coeffecient, electrical conductivity, power factor, and figure of merit of \textit{p}-type Co\textit{YZ} systems (\textit{Y} = V, Nb, Ta, and \textit{Z} = Si, Ge) at 1100 K, assuming $\tau$ = 10$^{-15}$ s. The reported values of CoTiSb at 850 K are taken as reference.}
\setlength{\arrayrulewidth}{0.5pt}
\begin{tabular*}{\textwidth}{c @{\extracolsep{\fill}} cccccc}
\hline
\hline
System		&\textit{n}   &S	    		       &$\sigma$			    &S$^2\sigma$			     &$ZT$  \\ 
            & (e/uc)	  &($\mu$V K$^{-1}$)	&(x 10$^{3}$ S cm$^{-1}$)	&$\mu$W cm$^{-1}$ K$^{-2}$	 &    \\	\hline		
CoTiSb \cite{Wu07}      &0.15   	  &153       	             &0.10          		&23.00            	         &0.45 \\ \hline     
CoVSi		&0.20		  &175			             &0.72				    &22.54		 		         &0.27 \\
CoNbSi		&0.18		  &163			             &0.99				    &26.67				         &0.74 \\
CoTaSi		&0.19		  &175			             &0.83				    &25.78		 		         &0.59 \\
CoVGe		&0.24		  &177			             &0.78				    &24.71		 		         &0.77 \\
CoNbGe		&0.20		  &173			             &0.93				    &28.15		 		         &0.64 \\
CoTaGe		&0.21		  &172			             &0.90				    &26.84		 		         &0.83 \\
\hline
\hline
\end{tabular*}
\label{zT}
\end{table*}

Interestingly, at 1100 K, all the proposed systems but CoVSi are found to have higher PF values on \textit{p}-type doping than the well-known CoTiSb; with maximum value for \textit{p}-type CoNbGe. The maximum reported PF value of \textit{p}-type CoTiSb is 23 $\mu$W cm$^{-1}$ K$^{-2}$ at 850 K \cite{Wu07}. This suggests the TE potential of the proposed systems at \textit{p}-type doping. Now, it becomes interesting to see how their thermal conductivity values will respond as the high $\kappa$ of hH alloys limits their applicability \cite{He16_II, Snyder08}. However, over the years, various approaches such as isoelectronic alloying, doping or nanostructuring have been helpful in reducing the $\kappa$ \cite{Snyder08, Dehkordi15}.

\begin{figure}
\centering\includegraphics[scale=0.45]{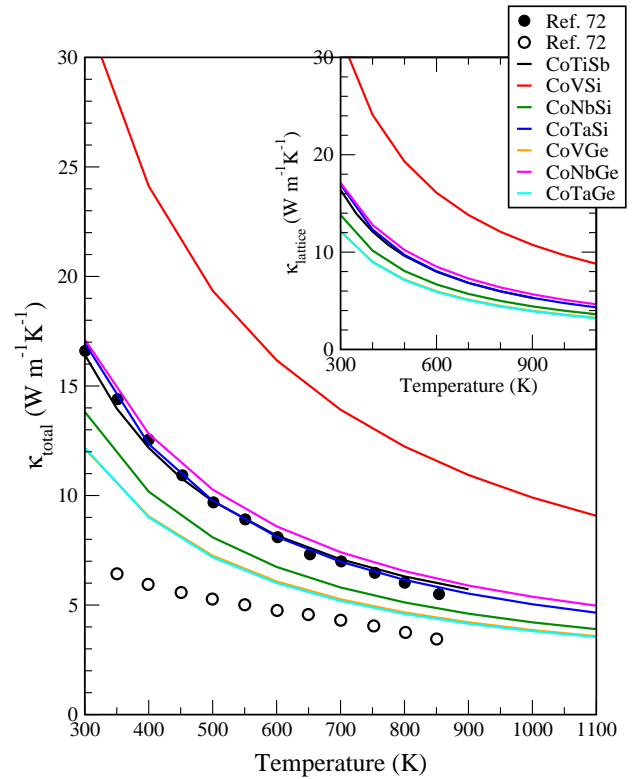}
\caption{The calculated thermal conductivity of Co\textit{YZ} systems (\textit{Y} = V, Nb, Ta, and \textit{Z} = Si, Ge) and
reference CoTiSb as a function of temperature by the single mode phonon relaxation time approximation (SMA). The inset of the
figure shows the lattice thermal conductivity as a function of temperature. The values of reported parent CoTiSb and 0.15 Fe-doped
CoTiSb are represented by filled and open circles, respectively.}
\label{kappa}
\end{figure}

Figure~\ref{kappa} shows the thermal conductivity as a function of the temperature of six cobalt-based systems, i.e., Si- and Ge-group. Thermal conductivity comprises two components, i.e., $\kappa_e$ and $\kappa_l$; $\kappa_e$ is calculated by Boltztrap code whereas $k_l$ is calculated by QE. The inset of the figure shows the lattice thermal conductivity. The contribution of $\kappa_e$ to the total $\kappa$ is almost negligible, as can be seen from the inset, and is well known for hH alloys \cite{Wu09, Kimura08}. Once again, we have taken the CoTiSb as the reference to corroborate our calculated values.  Note that the calculated $\kappa_l$ values are lowered by tenfold. This has been our observation from previous works that the calculated values of $\kappa_l$ by thermal2 code for ternary hH alloys are tenfold higher \cite{Zeeshan17_II, Zeeshan18}. Nevertheless, a nice agreement can be seen between our calculated and reported values of $\kappa$ of CoTiSb. We almost reproduced the reported $\kappa$ values of CoTiSb. This suggests the approximation used is quite reasonable and also substantiates our calculated values. 

Focussing on the other systems, the $\kappa$ of CoVSi is quite high whereas the CoTaGe has the lowest $\kappa$ values. This can be attributed to their effective masses. Heavy atoms can scatter the sound velocities effectively as compared to the lighter ones \cite{Chakraborty17}. Thus, the $\kappa$ of CoVSi is highest and that of the CoTaGe is lowest. In comparison to the well-known CoTiSb, the $\kappa$ of CoVSi is quite high whereas the $\kappa$ of CoNbGe slightly underestimate the reported $\kappa$ of CoTiSb. Where the $\kappa$ of CoTaSi almost parallels the CoTiSb, interestingly, the $\kappa$ of rest of the systems, i.e., CoNbSi, CoVGe, and CoTaGe, are lower than that of CoTiSb; with lowest $\kappa$ values for CoTaGe. The lower $\kappa$ of CoVGe, CoTaGe, and CoNbSi as compared to the CoTiSb suggests the TE potential of these systems in the proposed cubic symmetry. At 1100 K, the $\kappa$ of all the systems but CoVSi ranges 3.5--4.9 W m$^{-1}$ K$^{-1}$.


We have included the $\kappa$ of reported \textit{p}-doped CoTiSb to depict the importance of doping. The doping
lowers the $\kappa$ of CoTiSb almost by 2.4 times at 300 K and around 1.5 times at 850 K. This suggests the doping significantly
reduces the $\kappa$. As of now, owing to the computational challenges, we resort to the $\kappa$ of parent systems. To evaluate the $ZT$ of the considered systems, we have used PF of the doped systems whereas the $\kappa$ of the parent systems. As discussed, the $\kappa$ is known to reduce with doping, we assert that our calculated $ZT$ values are actually underestimated. 

Utilizing $\tau$ = 10$^{-15}$ s and $\kappa$ of the parent systems, the calculated values of $ZT$ at 1100 K are listed in Table~\ref{zT}. Also, Table~\ref{zT} shows the optimal doping level and the corresponding maximum power factor, Seebeck, and electrical conductivity of Si- and Ge-group. Once again, as a reference, the values of reported \textit{p}-type CoTiSb are also included. The calculated values of Seebeck coefficient range 163--177 $\mu$V K$^{-1}$; generally, the values of 150--250 $\mu$V K$^{-1}$ are considered to be a benchmark of a good TE material. The electrical conductivity values range 0.72--0.99 $\times$ 10$^3$ S cm$^{-1}$ whereas the carrier concentrations are of the order of 4--5 $\times$ 10$^{21}$ cm$^{-3}$; which is again a good number for a potential TE material. Further, the proposed doping levels are quite pragmatic and could be realized. The proposed optimal doping levels range from 0.18 to 0.24 hole doping per unit cell and could be achieved experimentally. For instance, the 0.18 hole doping in CoNbSi could be realized either by doping 18\% Si by Ga/In or by doping 18\% Nb by Zr/Hf or by substituting 18\% Co by Fe. Wu et al. reported high doping levels up to 42\% in CoTiSb, i.e., Co$_{0.572}$Fe$_{0.428}$TiSb \cite{Wu07}. Therefore, in light of experimental findings, we are optimistic that the proposed doping levels can be realized in the considered cobalt-based hH alloys. 

Importantly, the PF values of all the proposed systems are more competitive than the well-known CoTiSb. For the PF value of reported \textit{p}-type CoTiSb is 23 $\mu$W cm$^{-1}$ K$^{-2}$ at 850 K, the PF of all the systems but CoVSi on \textit{p}-type doping is higher than that of CoTiSb. The highest value was obtained in the case of CoNbGe, which is closely followed by other systems. Incorporating the $\kappa$ values of parent systems, as discussed, the highest $ZT$ obtained at 1100 K is 0.83 for CoTaGe, which is about twofold higher than the reported \textit{p}-type CoTiSb \cite{Wu07}. The $ZT$ value of CoVGe is 0.77, of CoNbSi is 0.74, of CoNbGe is 0.64, and that of CoTaSi is 0.74. The only system to have lower $ZT$ than that of reported CoTiSb is CoVSi, on account of its high thermal conductivity. All in all, except CoVSi, the $ZT$ values of all the proposed systems are higher than that of CoTiSb. This suggests the TE potential of all the proposed systems except CoVSi. As the dopants are not specified and we have incorporated $\kappa$ values of parents systems, our calculated values are actually a conservative estimate. We emphasize that the actual values of the figure of merit may even reach unity and will hopefully motivate the experimentalists to synthesize the proposed systems.

\subsection{Discussion and Conclusions}
Till now, the main focus in half-Heusler alloys has been centered on cubic symmetry, however still, there is a need for improvement of the figure of merit. Thus, a search of missing cubic 18-VEC half-Heusler alloys becomes essential. In this paper, we screened six cobalt-based 18-VEC ternary systems, i.e., CoVSi, CoNbSi, CoTaSi (Si-group) and CoVGe, CoNbGe, and CoTaGe (Ge-group), which crystallizes in orthorhombic symmetry. The idea was to somehow stabilize these systems in the cubic symmetry. Interestingly, we found that the three systems CoVGe, CoNbGe, and CoTaGe are energetically more favorable in the cubic symmetry -- favorable for thermoelectricity. Further, we discussed the importance of internal pressure and covalent interactions in stabilizing the CoVSi, CoNbSi, and CoTaSi in the cubic ground state. The pressure effect works by redistributing the structural and electronic properties in a system. It can bring the atoms close or move them away, as desired, leading to the change in the original symmetry. The various approaches discussed to introduce the internal pressure in the system includes the doping, insertion of nonreactive entities, and epitaxial strain. Every approach has its own merits and demerits. We emphasize more on the doping by heavier element to build the negative internal pressure as these alloys offer the privilege of vacant sites. However, these alloys are quite vulnerable to antisite disordering. 

Similarly, it is not relatively easier to introduce the nonreactive entities into the crystal structure. Considering the challenges for experimentalists, notwithstanding, the finding of Ge-group energetically more favorable in the cubic symmetry and their proposed transport properties can be seen as the real contribution of this work. On the other hand, the most intriguing finding is that the covalent interactions can stabilize these cobalt-based systems in the cubic symmetry. The principle of covalent interactions within the crystal lattice could be vital in speeding the search of missing cubic half-Heusler alloys. This principle can also be extended to stabilize the cubic symmetry of 18-VEC ternary alloys crystallizing in the non-cubic symmetry. Meanwhile, the transport properties of all the proposed systems but CoVSi are more impressive than the well-known CoTiSb. 

To conclude, utilizing \textit{ab initio} approach, we explored the orthorhombic to cubic phase transition and its impact on
thermoelectric properties in Co\textit{YZ} systems (\textit{Y} = V, Nb, Ta, and \textit{Z} = Si, Ge). It was found that the
Ge-group is energetically more favorable in the cubic symmetry whereas the Si-group can be stabilized in the cubic symmetry
by introducing internal pressure and covalent interactions within the crystal lattice. The transport properties of all the
proposed systems, except CoVSi, are more competitive than the well-known CoTiSb. The calculated $ZT$ value of \textit{p}-type CoTaGe at 1100 K is 0.83 which is twice as that of reported \textit{p}-type CoTiSb. The other impressive $ZT$ values obtained were 0.77, 0.74, 0.64, and 0.59 of \textit{p}-type CoVGe, CoNbSi, CoNbGe, and CoTaSi, respectively. We are optimistic that the actual values will be higher than what projected, as we have incorporated the $\kappa$ of undoped systems. We are confident that our findings would serve as a base for experimentalists to realize these cobalt-based systems in the proposed cubic symmetry. 

\subparagraph{Acknowledgments}
S.S. and M.Z. are thankful to MHRD and CSIR, respectively, for the support of a senior research fellowship. Computations were performed at IIT Roorkee, India and at IFW Dresden, Germany. We thank Ulrike Nitzsche for technical assistance.

\end{document}